\documentclass[12pt,a4paper]{article}
\usepackage{graphics}
\usepackage{amssymb}
\begin{document}
\begin{center} 
{\bf Evolution equations for pulse propagation in nonlinear media}
\vskip 0.5 cm
{\small Debabrata Pal, Amitava Choudhuri and B Talukdar }\\
{\small $^a$\it Department of Physics, Visva-Bharati University, Santiniketan 731235, India}
\vskip 0.2 cm
{\small \it e-mail : binoy123@bsnl.in}
\end{center}
\vskip 0.5 cm
\begin{abstract}
We show that the complex modified KdV (cmKdV) equation and generalized nonlinear Schr\"odinger (GNLS) equation belong to the  Ablowitz, Kaup, Newell and Segur or so-called AKNS hierarchy. Both equations do not follow from the action principle and are nonintegrable. By introducing some auxiliary fields we obtain the variational principle for them and study their canonical structures. We make use of a coupled amplitude-phase method to solve the equations analytically and derive conditions under which they can support bright and dark solitary wave solutions.
\end{abstract}
\vskip 0.5 cm
\vskip 1.0 cm
\noindent {\bf 1. Introduction}
\vskip 0.5 cm
\noindent In a pioneering work Zakharov and Shabat \cite{1} solved the nonlinear Schr\"odinger (NLS) equation by the use of inverse spectral method. Subsequently, Ablowitz, Kaup, Newell and Segur (AKNS) \cite{2} sought a generalization of the method and introduced the so-called AKNS hierarchy which involves a family of integrable equations associated with the Zakharov-Shabat eigenvalue problem. It is now fairly well known that many physically important integrable partial differential equations belong to the AKNS hierarchy. For example, besides the NLS equations, this hierarchy yields the KdV-, mKdV-, Sine-Gordon-, Harry-Dym equations as well as the constraint KP hierarchy \cite{3}. There are some relatively recent attempts  to disclose the possible connections between the two-component Camassa-Holm equation and AKNS hierarchy extended by a negative flow \cite{4,5}. Thus realization of physically important nonlinear evolution equations in the frame of AKNS model is still an interesting curiosity. \par In the present work we shall provide, in Sec. $2$, a derivation for two nonintgrable partial differential equations by using the integrable AKNS model for which there exists a well defined spectral problem. The equations of our interest are the complex modified KdV- (cmKdVI) \cite{6} and generalized third-order nonlinear Schrodinger (GNLS) equations \cite{7}. The cmKdVI equation describes the nonlinear steady-state propagation of lower-hybrid waves in a uniform plasma and more interestingly, its solutions reveal a close connection between classical soliton and envelope solitons. On the other hand, the GNLS equation is used to model propagation of ultra short pulses in optical fibers \cite{8} and is traditionally obtained from Maxwell's equations with special attention to nonlinear susceptibilities of the associated optical medium \cite{9}. \par An awkward analytical constraint for both cmKdVI- and GNLS equations is that these do not follow from the action principle to have a Lagrangian representation which plays a role in many applicative contexts \cite{10}. For an arbitrary differential equation one can always find the variational principle by taking recourse to the use of auxiliary field variables \cite{11}. We follow this route, in Sec. $3$, to construct expressions for Lagrangian densities of the cmKdV and GNLS equations and use them to study their canonical structures. In Sec. $4$ we make use of a coupled amplitude-phase formulation \cite{12} to solve these equations analytically and derive conditions under which cmKdVI- and GNLS equations can support bright and dark solitary wave solutions. Finally, in Sec. 5, we make some concluding remarks with a view to summarize our outlook on the present work.
\vskip 1.0 cm
\noindent {\bf 2. Derivation of cmKdVI- and GNLS equations using the AKNS hierarchy}
\vskip 0.5 cm
The AKNS hierarchy with the Zakharov-Shabat  eigenvalue problem is given by \cite{2}
$$i\left(\begin{array}{c}
u_t\\
-v_t
\end{array}\right)-\omega (2L_{zs})\left(\begin{array}{c}
u\\
v
\end{array}\right)=0\,\,.\eqno(1)$$
Here $u=u(x,\,t)$ and $v=v(x,\,t)$ represent two $(1+1)$ dimensional complex fields. The operator $\omega (2L_{zs})$ gives the dispersion relation of the linearization equation in the $u$ component with $L_{zs}$, an integrodifferential operator written as \cite{13}
$$L_{zs}={1\over {2i}}\left(\begin{array}{cc}
\frac{\partial}{\partial x}-2u\int_{-\infty}^xdy v&2u\int_{-\infty}^xdy u\\
-2v\int_{-\infty}^xdy v&-\frac{\partial}{\partial x}+2v\int_{-\infty}^xdy u
\end{array}\right)\,\,.\eqno(2)$$
The subscript ‘zs’ of $L$ merely indicates that the AKNS hierarchy is the family of integrable equations belonging to the Zakharov-Shabat  eigenvalue  problem. The NLS hierarchy can be obtained by choosing $\omega(2L_{zs})=(2L_{zs})^n\,\,,\,\,\,n=2,\,4,\,6\,...$ and $v=-u^*$. In particular, for $n=2$, $(1)$ gives the NLS equation. For $n=4$, we get the forth-order equation in the NLS hierarchy and so on. The choice for the dispersion relation amounts to demanding that $\omega(2L_{zs})$ is an entire function of the argument.
 \par To derive the nonintegrable equations of our interest we begin by introducing
 $$\omega(2L_{zs})=(-2L_{zs})^n\eqno(3)$$
  where $n=1,\,2,\,3\,...$. Note that for even values of $n$ the relation in $(3)$ coincides with the dispersion relation used for the NLS hierarchy. For $n=3$
$$\omega(2L_{zs})=-i\left(\begin{array}{cc}
a_{11}&a_{12}\\
a_{21}&a_{22}
\end{array}\right)\eqno(4)$$
with
$$a_{11}=\frac{\partial^3}{\partial x^3}-2u_{2x}\int_{-\infty}^xdy v-6u_xv-4uv\frac{\partial}{\partial x}+4u^2v\int_{-\infty}^xdy v+2u_x\int_{-\infty}^xdy v_y$$$$-2u\int_{-\infty}^xdy v_{2y}\,\,,\eqno(5a)$$
$$a_{12}=2u_{2x}\int_{-\infty}^xdy u+2uu_x-4u^2v\int_{-\infty}^xdy u+2u_x\int_{-\infty}^xdy v_y+2u\int_{-\infty}^xdy u_{2y}\,\,,\eqno(5b)$$
$$a_{21}=-2u_x\int_{-\infty}^xdy v_y-2vv_x-2v\int_{-\infty}^xdy v_{2y}+4uv^2\int_{-\infty}^xdy v-2v_{2x}\int_{-\infty}^xdy v\eqno(5c)$$
and
$$a_{22}=-\frac{\partial^3}{\partial x^3}+2v_{2x}\int_{-\infty}^xdy u+6v_xu+4uv\frac{\partial}{\partial x}-4uv^2\int_{-\infty}^xdy u-2v_x\int_{-\infty}^xdy u_y$$$$+2v\int_{-\infty}^xdy u_{2y}\,\,.\eqno(5d)$$
From $(1)$, $(3)$, $(4)$ and $(5)$ we get cmKdVI equation
$$u_t+2(|u|^2u)_x+u_{3x}=0\eqno(6)$$
for $v=-u*$. Similarly, for $n=5$ we get the fifth-order cmKdVI equation
$$ u_t-(8|u|^2u_{2x}+4|u|^4 u+2u^2u_{2x}^*+6u_x^2u^*)_x-10u_{x}^*u_{2x}+14u_{x}u_{2x}u^*-12uu_xu_{2x}^*-$$$$10u^2{u^*}^2u_x-u_{5x}=0\,\,.\eqno(7)$$
 The family of higher-order equations obtained in this way does not form a hierarchy in the same sense as used in the case of integrable equations.\par We have just seen that for $n=3$ the dispersion relation in $(3)$ generates cmKdV I equation. Let us now define
$$\omega(2L_{zs})=\sum_{n=2}^3(-2L_{zs})^n\,\,.\eqno(8)$$
Written explicitly $(8)$ becomes
$$\omega(2L_{zs})=-\left(\begin{array}{cc}
b_{11}+ia_{11}&b_{12}+ia_{12}\\
b_{21}+ia_{21}&b_{22}+ia_{22}
\end{array}\right)\eqno(9)$$
with
$$b_{11}=-\frac{\partial^2}{\partial x^2}+4uv+2u_{x}\int_{-\infty}^xdy v-2u\int_{-\infty}^xdy v_y\,\,,\eqno(10a)$$
$$b_{12}=-2u_x\int_{-\infty}^xdy u-2u\int_{-\infty}^xdy u_y\,\,,\eqno(10b)$$
$$b_{21}=-2v_x\int_{-\infty}^xdy v-2v\int_{-\infty}^xdy v_y\eqno(10c)$$
and
$$b_{22}=-\frac{\partial^2}{\partial x^2}+4uv+2v_{x}\int_{-\infty}^xdy u-2v\int_{-\infty}^xdy u_y\,\,.\eqno(10d)$$
From $(1)$, $(9)$ and $(10)$ we get GNLS equation
$$iu_t+2i|u|^2u_x+2i(|u|^2)_xu+u_{2x}+2|u|^2u+iu_{3x}=0\eqno(11)$$
for $v=-u^*$.
\vskip 1.0 cm
\noindent{\bf 3. Variational principle and canonical formulation}
\vskip 0.5 cm
The first step towards a canonical formulation of any system is to assure the existence of a Lagrangian. Both cmKdVI and GNLS equations in $(6)$ and $(11)$ invalidate the Helmholtz condition \cite{14} such that these equations are not Euler-Lagrange expressions. One standard method for finding variational principle \cite{11} for such equations is to introduce a set of auxiliary variables $v\{v^1,\,.......,v^i\}$ and consider the Lagrangian density ${\cal L}$
$$ {\cal L}(u,\,v)=v\Delta[u]\eqno(12)$$
 for the problem where $\Delta[u]=0$ defines an arbitrary system of differential equations with $\Delta=\Delta(\Delta_1,\,.....,\Delta_i)$. For the cmKdVI equation in $(6)$ we consider an additional complex field $v=v(x,t)$ and introduce the Lagrangian density
 $${\cal L}=v\left(u_t+4uu^*u_x+2u^2u_x^*+u_{3x}\right)-v^*\left(u^*_t+4uu^*u^*_x+2{u^*}^2u_x+u^*_{3x}\right)\eqno(13)$$
 to write the action principle as
 $$\delta\int {\cal L}\,dx\,dt=0\,\,.\eqno(14)$$
 Clearly, the Euler-Lagrange equations for $v$ and $v^*$ give the cmKdV I and its complex conjugate equations. The Euler-Lagrange equations for $u$ and $u^*$ yield the equations for $v$ and $v^*$. The set of four equations thus obtained can be written in the matrix form
{\small$$\left(\begin{array}{c}
u\\
u^*\\
v\\
v^*
\end{array}\right)_t=\left(\begin{array}{cccc}
-\partial_x^3-4|u|^2\partial_x&-2u^2\partial_x&0&0\\
-2{u^*}^2\partial_x&-\partial_x^3-4|u|^2\partial_x&0&0\\
0&0&-\partial_x^3-4|u|^2\partial_x&2{u^*}^2\partial_x\\ 
0&0&2{u}^2\partial_x&-\partial_x^3-4|u|^2\partial_x\end{array}\right)\times$$$$\left(\begin{array}{c}
u\\
u^*\\
v\\
v^*
\end{array}\right)\,\,.\eqno(15)$$}
The canonical momentum densities are
$$\pi=\frac{\partial{\cal L}}{\partial u_t}=v\,\,\,\,{\rm and}\,\,\,\,\pi^*=\frac{\partial{\cal L}}{\partial u^*_t}=-v^* \eqno(16)$$
corresponding to the Lagrangian density in $(13)$ which via Legendre transformation leads to the Hamiltonian density
$${\cal H}=v^*\left(4uu^*u^*_x+2{u^*}^2u_x+u^*_{3x}\right)-v\left(4uu^*u_x+2u^2u_x^*+u_{3x}\right)\eqno(17)$$
with the Hamiltonian written as 
$$H=\int {\cal H} dx\,\,.\eqno(18)$$
The equations in $(15)$ can be written in the Hamiltonian form
$$\dot\xi=\frac{\delta H}{\delta \eta}=\{\eta(x)\,,\,H(y)\}\eqno(19)$$
where $\xi$ and $\eta$ stand for appropriate field variables. It is easy to verify that $(19)$ is endowed with Poisson structures
$$\{u(x)\,,\,u(y)\}=\delta(x-y)\,\,,\,\,\,\{u(x)\,,\,u^*(y)\}=0\,\,,\,\,\,\{u(x)\,,\,v(y)\}=0\,\,,$$$$\,\,\,\{u(x)\,,\,v^*(y)\}=0\,\,,\eqno(20a)$$
$$\{u^*(x)\,,\,u(y)\}=0\,\,,\,\,\,\{u^*(x)\,,\,u^*(y)\}=\delta(x-y)\,\,,\,\,\,\{u(x)\,,\,v(y)\}=0\,\,,$$$$\,\,\,\{u(x)\,,\,v^*(y)\}=0\,\,,\eqno(20b)$$
$$\{v(x)\,,\,u(y)\}=0\,\,,\,\,\,\{v(x)\,,\,u^*(y)\}=0\,\,,\,\,\,\{v(x)\,,\,v(y)\}=\delta(x-y)\,\,,$$$$\,\,\,\{v(x)\,,\,v^*(y)\}=0\eqno(20c)$$
and
$$\{v^*(x)\,,\,u(y)\}=0\,\,,\,\,\,\{v^*(x)\,,\,u^*(y)\}=0\,\,,\,\,\,\{v^*(x)\,,\,v(y)\}=0\,\,,$$$$\,\,\,\{v^*(x)\,,\,v^*(y)\}=\delta(x-y)\,\,.\eqno(20d)$$
The Lagrangian and Hamiltonian densities for the GNLS equation in $(11)$ can be written as those in $(13)$ and $(17)$. The Lagrangian system of equations is given by
$$i\left(\begin{array}{c}
u\\
u^*\\
v\\
v^*
\end{array}\right)_t=\left(\begin{array}{cccc}
c_{11}&c_{12}&c_{13}&c_{14}\\
c_{21}&c_{22}&c_{23}&c_{24}\\
c_{31}&c_{32}&c_{33}&c_{34}\\
c_{41}&c_{42}&c_{43}&c_{44}
\end{array}\right)\left(\begin{array}{c}
u\\
u^*\\
v\\
v^*
\end{array}\right)\eqno(21)$$
with
$$c_{11}=c_{44}=-i\partial_x^3-\partial_x^2-4i|u|^2\partial_x\,\,,\eqno(22a)$$
$$c_{12}=-(2u^2+2iu^2\partial_x)\,\,,\eqno(22b)$$
$$c_{21}=-c_{12}^*\eqno(22c)$$
$$c_{22}=c_{33}=-i\partial_x^3+\partial_x^2-4i|u|^2\partial_x\,\,,\eqno(22d)$$
$$c_{13}=c_{14}=c_{23}=c_{24}=0\,\,,\eqno(22e)$$
$$c_{31}=2u^*v\,\,,\eqno(22f)$$
$$c_{32}=2uv\,\,,\eqno(22g)$$
$$c_{34}=2{u^*}^2-2i{u^*}^2\partial_x\,\,,\eqno(22h)$$
$$c_{41}=-c_{32}^*\,\,,\eqno(22i)$$
$$c_{42}=-c_{31}^*\eqno(22j)$$
and
$$c_{43}=-c_{34}^*\eqno(22k)$$
The set of equations in $(21)$ can be written in the Hamiltonian form as given in $(19)$ with Poisson structures similar to those in $(20)$
\vskip 1.0 cm
\noindent{\bf 4. Solitary wave solutions}
\vskip 0.5 cm
To obtain the solitary wave solutions of $(6)$ and $(11)$ we take recourse to the use of coupled amplitude-phase formulation \cite{12} and write $u(x,t)$ in the form
$$u(x,t) = P(\chi) exp[i(kx-\omega t)]\,\,, \,\,\,\chi =x + \beta t \,\,,\eqno(23)$$
with $\chi$ the travelling cordinate containing the group velocity $\beta$ of the wave packet.  Here the function $P$ is real. From $(6)$ and $(23)$ we obtain
$$\beta P_\chi+6P^2P_\chi+P_{\chi\chi\chi}-3k^2P_\chi+i(2kP^3-\omega P+3kP_{\chi\chi}-k^3P)=0\,\,.\eqno(24)$$
Equating the real and imaginary parts of $(24)$ seperately to zero we have 
$$P_{\chi\chi\chi}+6P^2P_\chi+(\beta-3k^2) P_\chi=0\eqno(25)$$
and
$$P_{\chi\chi}+{2\over3}P^3-\frac{\omega+k^3}{3k}P=0\,\,.\eqno(26)$$
The third order equation in $(25)$ can be integrated to write
$$P_{\chi\chi}+2P^3+(\beta-3k^2) P=0\,\,.\eqno(27)$$
Both equations $(26)$ and $(27)$ can be solved analytically. The solution of $(26)$ will depend explicitly on $\omega$ and $k$ while the solution of $(27)$ will have similar dependence on $\beta$ and $k$. In particular, $P(\chi)$'s will be given by
$$P(\chi)=(3k^2-\beta)^{1\over2}sech[(3k^2-\beta)^{1\over2}\chi]\eqno(26^\prime)$$
or
$$P(\chi)=(\frac{\omega+k^3}{k})^{1\over2}sech[(\frac{\omega+k^3}{3k})^{1\over2}\chi]\,\,.\eqno(27^\prime)$$
Here equations $(26^\prime)$ and $(27^\prime)$ refer to solutions of $(26)$ and $(27)$ when the first integrals of these equations are taken as zero. Clearly, the compatibility condition of these solutions implies that $(26^\prime)$ must satisfy $(27)$ and $(27^\prime)$ must satisfy $(26)$. This viewpoint yields the necessary and sufficient condition to get a relation among $\beta$, $\omega$ and $k$ and we have 
$$\beta=2 k^2-{\omega\over k}\,\,.\eqno(28)$$
Since the first integral of $(26)$ or $(27)$ gives the energy $E$ of the wave, we infer from $(26^\prime)$ and $(27^\prime)$ that the zero energy solution of the cmKdV equation represents a bright solitory wave solution provided the velocity $\beta$ satisfies the constraint in $(28)$. \par The choice $-\frac{(\beta-2k^2)^2}{4}$ and $-\frac{\omega+k^3}{4k}$ for the first integrals of $(26)$ and $(27)$ leads to the solutions
$$P({\chi})=\left(\frac{\beta-3k^2}{2}\right)^{1\over2}tanh\left[\left(\frac{\beta-3k^2}{2}\right)^{1\over2}(x+\beta t)\right]\eqno(26^{\prime\prime})$$
and
$$P({\chi})=\left(\frac{\omega+k^3}{2k}\right)^{1\over2}tanh\left[\left(\frac{\omega+k^3}{2k}\right)^{1\over2}(x+\beta t)\right]\eqno(27^{\prime\prime})$$
The compatibility condition of $(26^{\prime\prime})$ and $(27^{\prime\prime})$ gives the relation
$$\beta=4k^2+{\omega\over k}\,\,.\eqno(29)$$
It is of interest to note that for $\beta$ as given in $(29)$, the expressions for the first integral for $(26)$ and $(27)$ become identically equal. Thus we infer that for negative values of $E$ the cmKdVI equation supports dark solitory wave solution.\par We have carried out a similar analysis for the GNLS equation in $(11)$ and found that the zero energy solutions are given by
$$P({\chi})=\left(\frac{k^2-k^3-\omega}{1-\omega}\right)^{1\over2}sech\left[\left(\frac{k^2-k^3-\omega}{1-\omega}\right)^{1\over2}(x+\beta t)\right]\eqno(30)$$
or
$$P({\chi})=\left(3k^2-2k-\beta\right)^{1\over2}sech\left[\left(3k^2-2k-\beta\right)^{1\over2}(x+\beta t)\right]\eqno(31)$$
subject to the consistancy condition
$$\beta=k(3k-2)+\frac{k^2(k-1)}{1-\omega}+\frac{\omega}{1-\omega}\,\,.\eqno(32)$$
It is evident from $(30)$ or $(31)$, as with the cmKdVI equation, the zero energy solution of the GNLS equation also represents a bright solitory wave. Expectedly, the negative energy solution
$$P({\chi})=\left(\frac{\omega+k^3-k^2}{2(1-k)}\right)^{1\over2}tanh\left[\left(\frac{\omega+k^3-k^2}{2(1-3k)}\right)^{1\over2}(x+\beta t)\right]\eqno(33)$$
or
$$P({\chi})=\left(\frac{2k-\beta-3k^2}{2}\right)^{1\over2}tanh\left[\left(\frac{2k-\beta-3k^2}{2}\right)^{1\over2}(x+\beta t)\right]\eqno(34)$$
with
$$\beta=\frac{\omega}{1-k}-2k(1-k)\eqno(35)$$
corresponds to a dark solitory wave.
\vskip 1.0 cm
\noindent{\bf 5. Conclusions}
\vskip 0.5 cm
 We began by noting that a large number of physically important partial differential equation belongs to the AKNS hierarchy. This motivated us to examine if the nonintegrable cmKdVI and GNLS equations could be embedded into the same hierarchy. We provided derivation of these equations in the frame of the AKNS model. The equations are nonLagrangian even in the potential representation. But we found that one can obtain the variational principle for them by taking recourse to the use of some auxiliary fields. This provided a natural basis to study their canonical structures. We solved these equations by using a coupled amplitude-phase method and explicitly demonstrated the zero-energy solutions correspond to bright solitary waves while the negative-energy solutions represent dark solitary waves.

\end{document}